\documentclass{ws-ijmpa}

\begin{document}

\markboth{Lebed}{The $1/N_c$ Approach for Baryon Resonances}

\catchline{}{}{}{}{}

\title{The $1/N_c$ Approach for Baryon Resonances}

\author{Richard F. Lebed}
\address{Department of Physics \& Astronomy, Arizona State University,
Tempe, AZ 85287-1504, USA}

\maketitle

\begin{abstract}
I give a brief overview of the model-independent analysis of unstable
baryon resonances in the $1/N_c$ QCD expansion.  This approach
produces numerous surprising semi-quantitative phenomenological
predictions that are supported by available observations.
\end{abstract}


\section{Introduction}

This talk briefly summarizes a growing body of
work\cite{CL1,CLcompat,CDLN1,CLpent,CDLN2,CLSU3,CDLM} done with Tom
Cohen on unstable baryon resonances in the $1/N_c$ QCD expansion.
More in-depth synopses appear in Ref.~\refcite{talks}.

The $1/N_c$ expansion about the large $N_c$ QCD limit has been
extensively used to study baryons treated as stable states.  At $N_c
\! = \! 3$ the lowest-lying baryons fill a completely symmetric
spin-flavor SU(6) representation, the {\bf 56}.  Although the lightest
$N_c \! > \! 3$ baryons have $N_c$ valence quarks, the SU(6)
representation remains symmetric, as one shows\cite{DM} by imposing
order-by-order (in $1/N_c$) unitarity in $\pi N$ scattering ({\it
consistency conditions}).  Low-spin baryons in this multiplet
(analogues to $N$, $\Delta$) are split in mass at $O(1/N_c)$,\cite{J1}
meaning that such states are stable for large $N_c$.  Such analyses
use a Hamiltonian ${\cal H}$ with asymptotic baryon eigenstates, which
is expanded in SU(6)-breaking operators accompanied by powers of
$1/N_c$.

Baryon resonances, on the other hand, generically lie above the
ground-state multiplet by an $O(N_c^0)$ mass gap; since mesons like
$\pi$ have $O(N_c^0)$ masses, such states are generically unstable
against strong decay.  While treating excited multiplets like the
arbitrary-$N_c$ analogue of the SU(6) {\bf 70} using a Hamiltonian
approach has produced many interesting phenomenological results, such
a treatment for resonances is not entirely reasonable since it entails
grafting {\it ad hoc\/} sources of $q\bar q$ pairs (the
production/decay mechanism) onto ${\cal H}$, violating unitarity.
Moreover, resonances in such an ${\cal H}$ are not manifested as
excited states of stable baryons.

In fact, a resonance $R$ is properly represented by a complex-valued
pole in a scattering amplitude at position $z_R \! = \! M_R -
i\Gamma_R/2$.  Such an approach can be developed using tools originally
appearing in the context of chiral soliton models, for such models
naturally accommodate baryon resonances as excitations resulting from
scattering of mesons off ground-state baryons.  They are consistent
with the large $N_c$ limit because the solitons are heavy,
semiclassical objects compared to the mesons.  In contrast to quark
models, chiral soliton models tend to fall short in providing detailed
spectroscopy and decay parameters for baryon resonances, particularly
at higher energies.  Nevertheless, one may use the scattering or
soliton {\em picture\/} (no specific model) to study baryon resonances
as well as the full scattering amplitudes in which they appear, and
relate it for $N_c$ large to the Hamiltonian or quark {\em picture}.

\section{Amplitude Relations} \label{amp}

The relevant physical observables in real data are scattering
amplitudes, in which resonances appear as complex poles.  The role of
large $N_c$ is to treat the ground-state baryon, off which mesons
scatter, as a heavy field manifesting the consistency conditions.
These conditions turn out to impose\cite{KapSavMan} t-channel
constraints: Amplitudes with $|I_t \! - \! J_t| \! =
\! n$ first appear at order (both relative and absolute) $N_c^{-n}$,
leading to relations among the amplitudes holding at various orders in
$1/N_c$.  The t-channel expressions may be reexpressed in the s
channel using group-theoretical crossing relations, in which the $I_t
\! = \! J_t$ rule becomes\cite{MM} the statement that the underlying
dynamical degrees of freedom are {\it reduced amplitudes\/} $s$
labeled by $K$, the eigenvalue of ${\bf K} \! \equiv \! {\bf I} \! +
\! {\bf J}$.  The $\pi N$ and $\eta N$ partial waves of orbital
angular momentum $L$ are

\begin{equation}
L^\pi_{IJ} = 2 \sum_K (2K+1)
\left\{ \begin{array}{ccc} K & I & J \\
\frac 1 2 & L & 1 \end{array} \right\}^2
s^\pi_{K L L} \ , \ \ \label{MPeqn1}
L^\eta_{IJ} = s_{LLL}^\eta \ ,
\end{equation}
respectively, which are special cases of an
expression\cite{CLcompat,Mat9j} that holds for ground-state baryons of
arbitrary spin (= isospin) and mesons of arbitrary spin and isospin.

Expressions like (\ref{MPeqn1}) are useful because full observable
partial waves outnumber reduced amplitudes $s$, giving rise to a
number of linear relations among the measured amplitudes that hold at
leading [$O(N_c^0)$] order.  A pole in one partial wave implies a pole
in one of the $s$, which in turn contributes to several others partial
waves.  A single pole therefore generates a multiplet of baryon
resonances degenerate in mass and width.\cite{CL1} Moreover, the
existence of an s-channel pole in an intermediate state cannot depend
upon the states used to form it (except through conservation of
quantum numbers), and so the resonant poles are labeled uniquely by
$K$.

For example, Eqs.~(\ref{MPeqn1}) applied to $I \! = \! \frac 1 2$,
$J^P \! = \! {\frac 1 2}^-$ states ($N_{1/2}$) gives $S^\pi_{11} \! =
\!  s^\pi_{100}$ and $S^\eta_{11} \! = \! s^\eta_0$, while in the $I
\! = \! \frac 1 2$, $J^P \! = \! {\frac 3 2}^-$ ($N_{3/2}$) channel,
$D^\pi_{13} \! = \! \frac 1 2 ( s^\pi_{122} \! + \! s^\pi_{222} )$
(parity being determined by $L$).  A pole in $S_{11}^\pi$ ($K \!  =
\! 1$) implies a degenerate pole in $D_{13}^\pi$, because the latter
contains a $K \! = \!  1$ amplitude.  Clearly, degenerate towers of
resonances develop with $K \! = \! 0, 1, 2$, {\it etc.}  On the other
hand, one type of $N_{1/2}$ pole ($K \! = \!  1$) couples
predominantly to $\pi$ and the other ($K \! = \! 0$) to $\eta$.

In comparison, the quark-picture large $N_c$ analogue of the excited
multiplet $({\bf 70}, 1^- )$ has 2 $N_{1/2}$ and 2 $N_{3/2}$ states,
just as for $N_c \! = \! 3$.  One can show that the numerous
nonstrange states in this multiplet have only 3 distinct mass
eigenvalues out to $O(N_c^0)$,\cite{CL1,CCGL,PS} which we label
$m_{0,1,2}$.  Comparing the quantum numbers and multiplicities of such
states with the results of Eqs.~(\ref{MPeqn1}), one immediately
finds\cite{CL1} that exactly the required resonant poles are obtained
if each $K$ amplitude, $K \! = \! 0,1,2$, contains precisely one pole,
located at the value $m_K$.  This quark-picture multiplet of excited
baryons is found to be {\em compatible\/} with, {\it i.e.}, consist of
a complete set of, multiplets classified by $K$.

One can prove\cite{CLcompat} compatibility for all nonstrange baryon
multiplets in the SU(6)$\times$O(3) shell picture, as well as for the
whole $({\bf 70}, 1^- )$ multiplet\cite{CLSU3} if $K \! = \!
\frac 1 2, \frac 3 2$ poles are added.  It is important to note that
compatibility does {\em not\/} imply SU(6) is an exact symmetry at
large $N_c$ for resonances as it is for ground states; rather, it says
that SU(6)$\times$O(3) multiplets are complete but {\em reducible\/}
at large $N_c$.  In the above example, $m_{0,1,2}$ are split at
$O(N_c^0)$ and therefore label distinct multiplets.  Indeed, large
$N_c$ by itself does not mandate the existence of {\em any\/} baryon
resonances, but it does say that if even one exists, it must be a
member of a well-defined multiplet.

\section{Phenomenology} \label{phenom}

It is possible to obtain testable predictions for the decay channels,
even using just the leading order in $1/N_c$.  For example, we noted
that the $K \! = \! 0(1)$ $N_{1/2}$ resonance couples only to
$\eta$($\pi$).  Indeed, the $N(1535)$ resonance decays to $\eta N$
with a 30--55$\%$ branching ratio (BR) despite lying barely above that
threshold, while the $N(1650)$ decays to $\eta N$ with only a
3--10$\%$ BR despite having much more comparable $\pi N$ and $\eta N$
phase spaces.  This pattern clearly suggests\cite{CL1} that the
$\pi$-phobic $N(1535)$ should be identified with $K \! = \!  0$ and
the $\eta$-phobic $N(1650)$ with $K \! = \! 1$, the first fully field
theory-based explanation for these observations.  Similarly, the
$\Lambda(1670)$ lies only 5~MeV above the $\eta\Lambda$ threshold but
has a 10--25$\%$ BR to this channel, and thus appears to lie in a $K
\! = \! 0$ SU(3) octet with the $N(1535)$.\cite{CLSU3} Large $N_c$
also provides\cite{CLSU3} SU(3) selection rules: Octet resonances
preferentially produce $\pi$ and $\eta$ in their decays, while singlet
$\Lambda$'s produce $\overline{K}$'s.

Configuration mixing between states with 1, 2, {\it etc.} excited
quarks, which requires an {\it ad hoc\/} treatment in the quark
picture, naturally occurs in the scattering picture.  If one finds two
resonances of comparable mass and identical quantum numbers, and one
has a natural broad $O(N_c^0)$ width, can the other have a narrow
$O(1/N_c)$ width?  Unless a conservation law forbids the mixing, both
are expected to have $\Gamma \! = \! O(N_c^0)$.  Indeed, one can
exhibit\cite{CDLN1} $O(N_c^0)$ operators that allow such mixing.

The mandate that resonances occur in multiplets applies not only to
conventional but exotic baryons as well.  For example, a state with
the quantum numbers of the controversial $\Theta^+(1540)$ ($I \! = \!
0$, $J \! = \! \frac 1 2$) would necessarily have
partners\cite{CLpent} with $I \!  = \! 1$, $J \! = \! \frac 1 2, \frac
3 2$ and $I \! = \!  2$, $J \! = \! \frac 3 2, \frac 5 2$, degenerate
in mass and width up to $O(1/N_c)$ corrections.  Indeed, such a
$\Theta^+$ in an SU(3) $\overline{\bf 10}$ would have degenerate {\bf
27} and {\bf 35} partners as well,\cite{CLpent} an enumeration
matching that found in a quark picture calculation.\cite{JMpent}

First-order $1/N_c$ corrections in amplitude relations like
Eqs.~(\ref{MPeqn1}) are handled\cite{CDLN2} (at least in the
nonstrange case) simply by including $|I_t \! - \! J_t| \! = \! 1$
amplitudes in the analysis and identifying linear relations surviving
among the enlarged set of physical amplitudes.  When confronted with
data, relations obtained in this way are typically a factor $N_c
\! = \! 3$ better satisfied than those that only hold to $O(N_c^0)$.

Processes other than meson-baryon scattering can also be treated using
the scattering approach, given only the $I$ and $J$ quantum numbers of
the field coupling to the baryon along with the corresponding $N_c$
power suppression of each coupling.  For example, each multipole
amplitude in pion photoproduction can be expressed in this
way,\cite{CDLM} once the photon is divided into its $I \! = \! 1$ and
$I \! = \! 0$ pieces.  Amplitude relations thus derived, including
$1/N_c$ corrections, can be compared to data.  The comparisons are
favorable except in the resonant region, where the effects of finite
$1/N_c$ splittings between resonances degenerate in the $N_c \! \to \!
\infty$ limit obscure agreement.  In fact, the helicity amplitudes on
corresponding resonances derived from such seemingly disparate
amplitudes turn out to agree remarkably well with $1/N_c$
expectations.\cite{CDLM}

The most interesting unsolved problems using the scattering method
involve a proper treatment of $1/N_c$ effects.  First, the $I_t \! =
\! J_t$ and its incorporation of $1/N_c$-suppressed amplitudes must be
generalized to strange resonances.  Second, many spurious states arise
when $N_c \! > \! 3$; explaining how they decouple when $N_c$ is tuned
to 3 will open a broad new front for phenomenological results.
Finally, a marriage of this approach with chiral symmetry
(straightforward because of the connection to soliton models) will
provide a very powerful additional set of physical constraints.

\section*{Acknowledgements}

The author's research was supported by the N.S.F\@. through grant
PHY-0140362.


\begin{thebibliography}{99}

\bibitem{CL1}
T.D.~Cohen and R.F.~Lebed, {\it Phys.\ Rev.\ Lett.} {\bf 91}, 012001
(2003); {\it Phys.\ Rev.} D {\bf 67}, 096008 (2003).

\bibitem{CLcompat}
T.D.~Cohen and R.F.~Lebed, {\it Phys.\ Rev.} D {\bf 68}, 056003 (2003).

\bibitem{CDLN1}
T.D.~Cohen, D.C.~Dakin, A.~Nellore, and R.F.~Lebed, {\it Phys.\ Rev.}
D {\bf 69}, 056001 (2004).

\bibitem{CLpent}
T.D.~Cohen and R.F.~Lebed, {\it Phys.\ Lett.} B {\bf 578}, 150 (2004);
{\it ibid.} {\bf 619}, 115 (2005).

\bibitem{CDLN2}
T.D.~Cohen, D.C.~Dakin, A.~Nellore, and R.F.~Lebed, {\it Phys.\ Rev.}
D {\bf 70}, 056004 (2004).

\bibitem{CLSU3}
T.D.~Cohen and R.F.~Lebed, {\it Phys.\ Rev.} D {\bf 70}, 096015
(2004); {\tt hep-ph/0507267} (to appear in {\it Phys.\ Rev.} D).

\bibitem{CDLM}
T.D.~Cohen, D.C.~Dakin, R.F.~Lebed, and D.R.~Martin, {\it Phys.\ Rev.}
D {\bf 71}, 076010 (2005).

\bibitem{talks}
R.F.~Lebed, {\tt hep-ph/0406236}, in {\it Continuous Advances in QCD
2004}, ed.\ by T.~Gherghetta, World Scientific, Singapore (2004); {\tt
hep-ph/0501021}, in {\it Large $N_c$ QCD 2004}, ed.\ by J.L.~Goity
{\it et al.}, World Scientific, Singapore (2005); T.D.~Cohen, {\it
Nucl.\ Phys.} A {\bf 755}, 40 (2005).

\bibitem{DM}
R.F.~Dashen and A.V.~Manohar, {\it Phys.\ Lett.} B {\bf 315}, 425
(1993).

\bibitem{J1} E. Jenkins, Phys.\ Lett.\ B {\bf 315}, 441 (1993).

\bibitem{KapSavMan}
D.B.~Kaplan and M.J.~Savage, {\it Phys.\ Lett.} B {\bf 365}, 244
(1996); D.B.~Kaplan and A.V.~Manohar, {\it Phys.\ Rev.} C {\bf 56}, 76
(1997).

\bibitem{MM}
M.P.~Mattis and M.~Mukerjee, {\it Phys.\ Rev.\ Lett.} {\bf 61}, 1344
(1988).

\bibitem{Mat9j}
M.P.~Mattis, {\it Phys.\ Rev.\ Lett.} {\bf 56}, 1103 (1986).

\bibitem{CCGL}
C.E.~Carlson, C.D.~Carone, J.L.~Goity, and R.F.~Lebed, {\it Phys.\
Lett.} B {\bf 438}, 327 (1998); {\it Phys.\ Rev.} D {\bf 59}, 114008
(1999).

\bibitem{PS}
D.~Pirjol and C.~Schat, {\it Phys.\ Rev.} D {\bf 67}, 096009 (2003).

\bibitem{JMpent}
E.~Jenkins and A.V.~Manohar, {\it Phys.\ Rev.\ Lett.} {\bf 93}, 022001
(2004); {\it JHEP} {\bf 0406}, 039 (2004).

\end{thebibliography}
\end{document}